\documentclass[letterpaper,12pt,oneside,onecolumn]{article}%
\usepackage{geometry}%
\usepackage{titlesec}%
\titleformat{\section}[hang]{\normalfont\normalsize\bfseries}{\thesection}{12pt}{\centering}%
\titleformat{\subsection}[display]{\normalfont\normalsize}{\thesubsection}{12pt}{\underline}%
\titleformat{\subsubsection}[runin]{\normalfont\normalsize}{\thesubsubsection}{12pt}{\underline}%
\newcommand{\PaperTitle}[1]{%
\begin{center}%
    \begin{large}%
        \textbf {#1} \\%
    \end{large}%
\end{center}%
}%
\newcommand{\AuthorList}[1]{%
\begin{center}%
    {#1} \\%
\end{center}%
}%
\newcommand{\AuthorAffiliations}[1]{%
\begin{center}%
    {#1} \\%
\end{center}%
}%
\newcommand{\Keywords}[1]{%
\begin{center}%
   Keywords: {#1} \\%
\end{center}%
}%

\usepackage[pdftex]{graphicx}
\usepackage[utf8]{inputenc}
\usepackage{subfigure}
\usepackage{amsmath,amssymb}
\usepackage{bm}	
\usepackage{xspace} 
\usepackage[pdftex]{hyperref}
\usepackage{cite}

\begin{document}

\PaperTitle{First order pyramidal slip \\
of $1/3\,\langle1\bar{2}10\rangle$ screw dislocations in zirconium\footnote{Article
published in Metall. Mater. Trans. A, 
symposium \emph{Multiscale perspectives on plasticity in hcp metals}, 
TMS 2014 annual meeting;
\url{http://dx.doi.org/10.1007/s11661-014-2568-7}}}

\AuthorList{Nermine Chaari$^1$, Emmanuel Clouet$^1$, and David Rodney$^2$}%

\AuthorAffiliations{$^1$CEA, DEN, Service de Recherches de Métallurgie physique;\\
Gif-sur-Yvette 91191, France\\
$^2$Institut Lumière Matière, Université Lyon 1, CNRS, UMR 5306;\\
Villeurbanne, 69622, France}%


\section{Abstract} 

Atomistic simulations, based either on an empirical interatomic potential or on ab initio calculations, are used to study the pyramidal glide of a $1/3\ \langle1\bar{2}10\rangle$ screw dislocation in hexagonal close-packed zirconium. Generalized stacking fault calculations reveal a metastable stacking fault in the first order pyramidal $\lbrace 10\bar{1}1 \rbrace$ plane, which corresponds to an elementary pyramidal twin. This fault is at the origin of a metastable configuration of the screw dislocation in zirconium, which spontaneously appears when the dislocation glides in the pyramidal plane.

\section{Introduction} 

Hexagonal close-packed (hcp) Zirconium is an important material for the nuclear industry where it is used as structural component in nuclear reactors. In particular the cladding of nuclear fuel is made of zirconium alloys. Like most crystalline material, the mechanical behavior is mainly driven by dislocations motion. 

In $\alpha$-zirconium, dislocations with Burgers vector $\vec{a} = 1/3 \ [1\bar{2}10]$, named $\langle a\rangle$ dislocations, are the most frequently observed with transmission electron microscopy \cite{Rapperport1959, Caillard2003, Ferrer2000}. These dislocations glide principally in prismatic $\lbrace 10\bar{1}0 \rbrace$ planes \cite{Vitek2008, Caillard2003, Kubin2013} due to a lower critical resolved shear stress than in the basal and pyramidal planes \cite{Rapperport1959, Akhtar1971, Akhtar1975b, Tyson1967, Akhtar1975c, Ferrer2000}. 
At low temperature, screw components of $\langle a \rangle$ dislocations can be distinguished as long rectilinear  segments while mixed and edge components are observed in their equilibrium state as curved lines.  
This is because screw dislocations have a larger lattice friction opposing  their motion and making them less mobile compared to mixed and edge dislocations \cite{Rapperport1959, Caillard2003, Ferrer2000}. For this reason, screw dislocations with Burgers vector $\vec{a}$ control the material plasticity at low temperature and are mostly considered in the literature. In addition, experiments show that the ease of glide of $\langle a\rangle$ screw dislocations in the prismatic planes is strongly temperature dependent and also decreases when the amount of impurities such as oxygen, sulfur and carbon, increases in the material \cite{Kubin2013, Baldwin1968, Sastry1971, Mills1968, Ferrer2000}. 

At higher temperatures and strain levels, secondary slip systems are activated such as $1/3\ \langle 1\bar{2}13 \rangle $ first order pyramidal slip \cite{Numakura1991}, which has been evidenced at room temperature as an important glide system to accommodate the crystal  deformation along the $\langle 0001 \rangle $ direction. Thermal activation also enhances $\langle a\rangle$ dislocation cross-slip. Experimental evidence shows that above 300\,K (573$^{\circ}$C), screw dislocations with Burgers vector $\vec{a}= 1/3 \ [1\bar{2}10]$ initially gliding in the prismatic planes, may leave their habit plane to glide in a first order pyramidal  $\lbrace 10\bar{1}1 \rbrace$ plane ($\pi_1$)\cite{Baldwin1968, Ferrer2000, Rautenberg2012}, or less frequently in a basal $\lbrace 0001 \rbrace$ plane \cite{Akhtar1971, Akhtar1973, Yapici2009, Knezevic2013}. Screw $\langle a\rangle$ dislocations cross slip is more frequently observed with increasing impurity content, especially with oxygen, while the hardening effect due to impurities manifests itself on the prismatic glide \cite{Ferrer2000, Sastry1971, Baldwin1968, Kubin2013}, as mentioned above. The same dislocation behavior has also been evidenced in titanium \cite{Churchman1954, Shechtman1973, Rosi1953, Farenc1993, Naka1983, Naka1988}, a transition metal with similar properties to zirconium.

In agreement with the experiments, atomistic simulations have established that, in pure zirconium, a screw dislocation with Burgers vector $\vec{a}$ dissociates spontaneously in the prismatic plane into two partial dislocations with Burgers vector $\vec{a}/2$ \cite{clouet2012, Bacon2002}. The dissociation is explained by a stable stacking fault with low energy in the prismatic plane \cite{Domain2004a, Poty2011, clouet2012}. In turn, dissociation leads to a low lattice friction that makes screw dislocations glide easily in the prismatic planes \cite{Khater2010, clouet2012}.  
 Considering pyramidal and basal slip of $\langle a\rangle$ dislocations in pure zirconium, it has been shown in a previous article that both slip systems share the same thermally-activated mechanism involving a metastable core structure, where the screw dislocation with Burgers vector $\vec{a}$ partially spreads in a pyramidal plane \cite{Chaari2014}. The aim of the present paper is to study in more details pyramidal slip and to justify the glide mechanism suggested in Ref. \cite{Chaari2014}.     
  
As dislocation mobility results from their core structure, accurate atomic-scale study of the dislocation core is required to understand their mobility. Besides, it has been shown that in hcp transition metals, the relative ease of dislocation glide is directly related to the stacking fault energies in the glide planes, which are in turn controlled by the electronic structure of the metal \cite{Legrand1984}. Atomistic simulations incorporating a full description of the electronic structure are therefore necessary to model dislocations in zirconium. In the present work, we used both ab initio calculations and an empirical potential to calculate the energy associated with shearing the hcp lattice in a $\pi_1$ pyramidal plane, both homogeneously to determine generalized stacking faults, and inhomogeneously when a $\langle a \rangle$ dislocation, initially dissociated in a prismatic plane glides along a $\pi_1$ pyramidal plane.
  
\section{Methods}  

  As described in previous papers \cite{clouet2012,Chaari2014}, we performed ab initio calculations based on the density functional theory using the PWSCF code \cite{Giannozzi2009}. Atomistic simulations with the embedded atom method (EAM) potential developed by Mendelev and Ackland (potential \#3)\cite{Mendelev2007} were also performed to study the effect of simulation cell size.  A comparison between results obtained with both energy models was established to assess the ability of this empirical potential to describe pyramidal slip.
  
   Dislocation core structure and glide are directly related to the stacking fault energy in the glide plane. To characterize the shearing of the crystal in the pyramidal plane, we calculated the generalized stacking fault \cite{Vitek2008} in this plane using fully periodic boundary conditions. Only one fault is introduced in the simulation cell, and the corresponding shift is applied to the periodicity vector perpendicular to the fault plane.
Atom relaxation is only allowed in the direction perpendicular to the fault plane, but we also performed calculations with full atomic relaxation to identify stable stacking faults.  
 
  To model dislocations, we used a fully periodic arrangement of dislocation dipole described as an S arrangement in Ref. \cite{clouet2012}. The dimensions of the simulation cells are $n \times \sqrt{3} \times a$ in the $[10\bar{1}0]$ direction, $m \times c$ in the $[0001]$ direction and $a$ in the $[1\bar{2}10]$ direction along the dislocation line ($n$ and $m$ are two integers). Atoms are relaxed until all components of the atomic forces are smaller than 10 meV/\AA. We checked for some configurations that relaxation at 2 meV/\AA \ does not change any of our results.
  We employed the Nudged elastic Band (NEB) method \cite{Henkelman2000} to determine the energy barrier against dislocation glide in the first order pyramidal plane. This method gives the minimum energy path between two stable states. All  energy barriers are calculated while moving both dislocations composing the dipole in the same direction and the path is relaxed with a tolerance on atomic forces of 20 meV/\AA . 
  
\begin{figure}[!tbh]
\begin{center}
  \includegraphics[width=0.8 \linewidth]{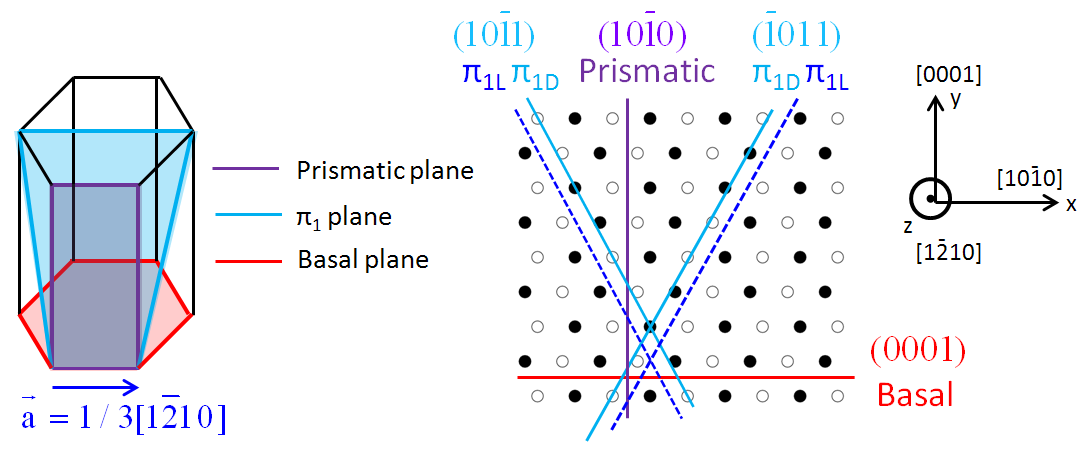}
\caption{Hexagonal close-packed structure showing the different potential glide planes for a screw dislocation of Burgers vector $\vec{a}=1/3\ [1\bar{2}10] $. A projection perpendicular to $\vec{a}$ is shown on the right, where atoms are sketched by circles with a color depending on their (1$\bar{2}$10) prismatic plane. (color online)}
\label{figure1}
\end{center}
\end{figure}  
 
\section{Stacking fault energy in the pyramidal plane}

Since dislocation glide is directly related to the stacking fault energy in the corresponding plane, we started by investigating stacking fault energies in the first order pyramidal plane. In the hcp structure, pyramidal planes are corrugated. As a consequence, there are two different ways to shear the crystal along a pyramidal plane. The crystal might be sheared either inside a corrugated pyramidal plane, which we call a dense plane $\pi_{1D}$, or between two pyramidal corrugated planes, i.e. inside  a loose plane $\pi_{1L}$ (Fig. \ref{figure1}). 
The generalized stacking fault is calculated for both types of pyramidal planes using both ab initio calculations and the EAM potential.

\begin{figure}[!tbh]
\begin{center}
\hfill
	\subfigure [$\pi_{1L}$] {\includegraphics[width=0.36 \linewidth]{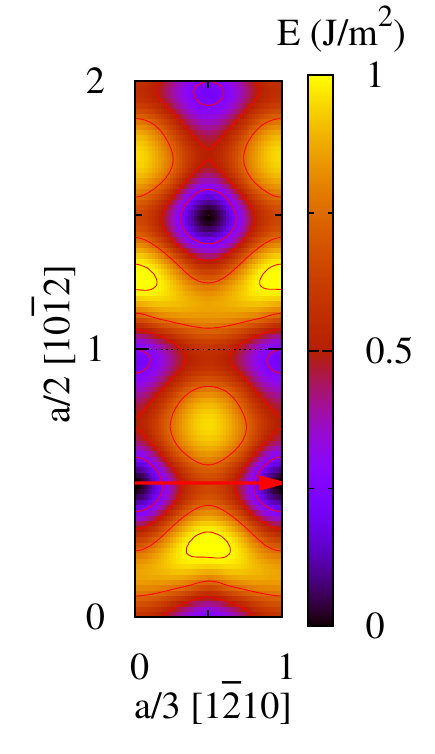}}
	\hfill
	\subfigure [$\pi_{1D}$] {\includegraphics[width=0.36 \linewidth]{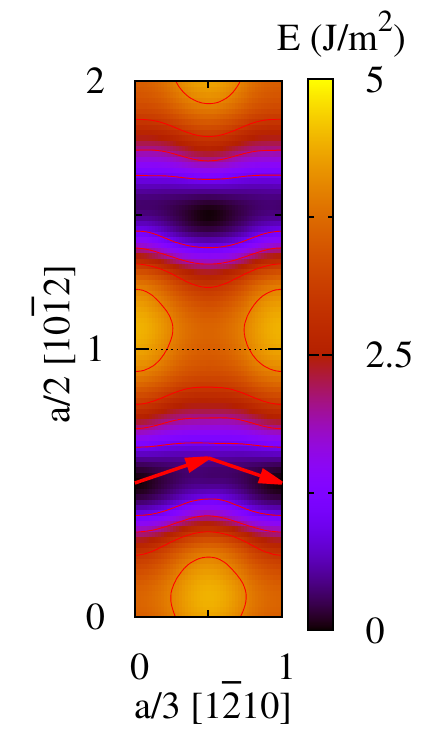}}
\hfill
\hfill
\caption{Generalized staking fault energy in the first order pyramidal plane $(10\bar{1}1)$ calculated with ab initio.
The crystal is sheared either (a) in the loose plane $\pi_{1L}$, 
or (b) in the dense plane $\pi_{1D}$. The red arrows show the fault vector $a/3 \ [1\bar{2}10]$ in (a) and its decomposition in (b) corresponding to the metastable stacking fault obtained after a full relaxation of atoms in the $\pi_{1D}$ plane. (color online)}
\label{projections}
\end{center}
\end{figure}

 We used a simulation cell with a height $h= q\times \zeta$, where $\zeta \sim 5.9$\,{\AA} is the height of the elementary cell and $q=4$ is the number of the atomic planes separating two faults. The convergence of our results with respect to the  simulation cell height has been checked with the EAM potential.
 
The results obtained with ab initio calculations and with the EAM potential in both pyramidal $\pi_{1D}$ and $\pi_{1L}$ planes are in good agreement. We show in Fig. \ref{projections} the $\gamma$-surfaces obtained with ab initio calculations in both pyramidal planes. 
From a general point of view, the $\gamma$-surfaces show that shearing the crystal in a $\pi_{1D}$ plane costs higher energy than in the $\pi_{1L}$ plane. This is a consequence of the fact that atoms are close to each other in this plane, and the shearing may bring them closer, which strongly increases the energy. But this high energy landscape is explored only when a $[10\bar{1}2]$ fault component is involved. Focusing now on the $[1\bar{2}10]$ direction, which is the relevant direction for $\langle a \rangle$ dislocation glide, the $\gamma$-surfaces calculated in both $\pi_{1L}$ and $\pi_{1D}$ plane show a valley of low energy along this direction.  
 
\begin{figure}[!tbh]
\begin{center}
	\subfigure [$\pi_{1L}$] {\includegraphics[width=0.49 \linewidth]{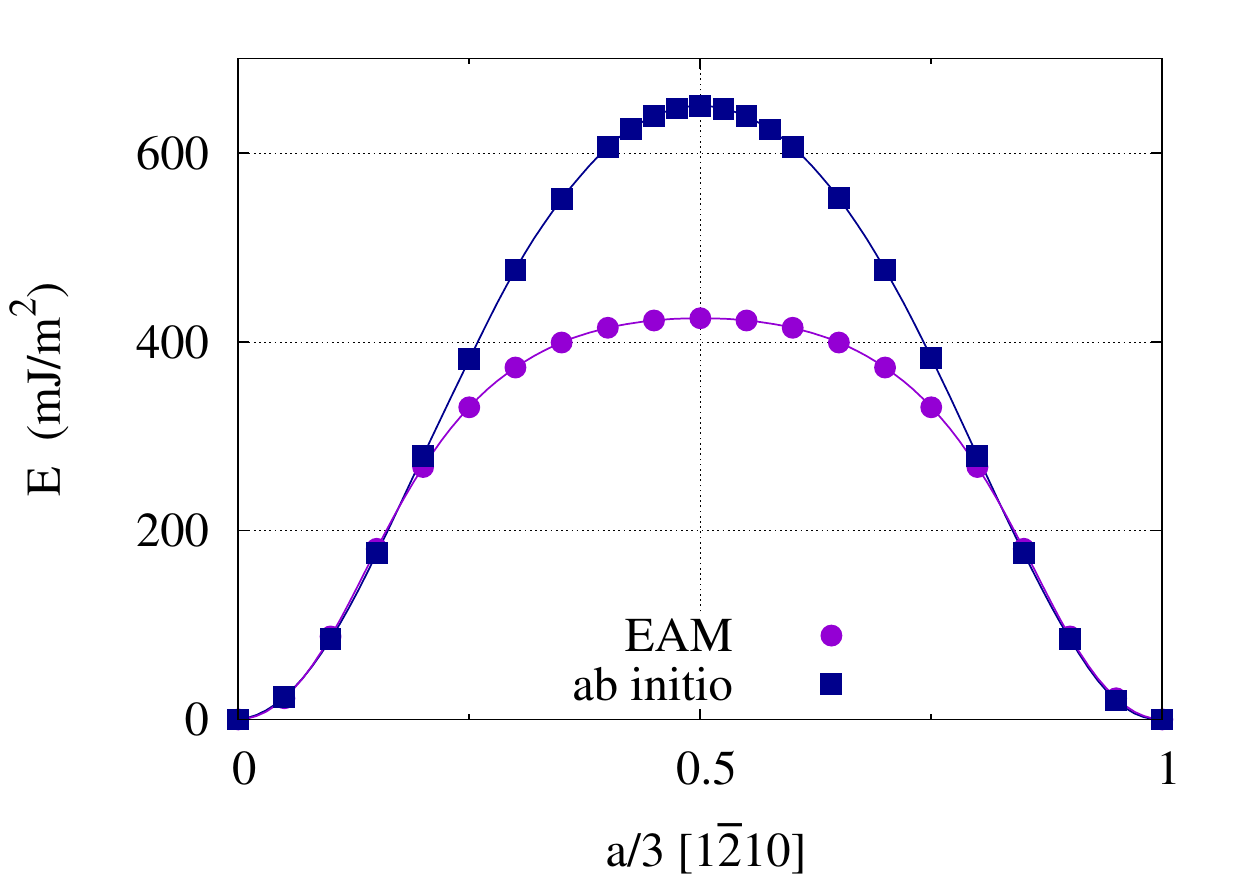}}
	\hfill
	\subfigure [$\pi_{1D}$] {\includegraphics[width=0.49 \linewidth]{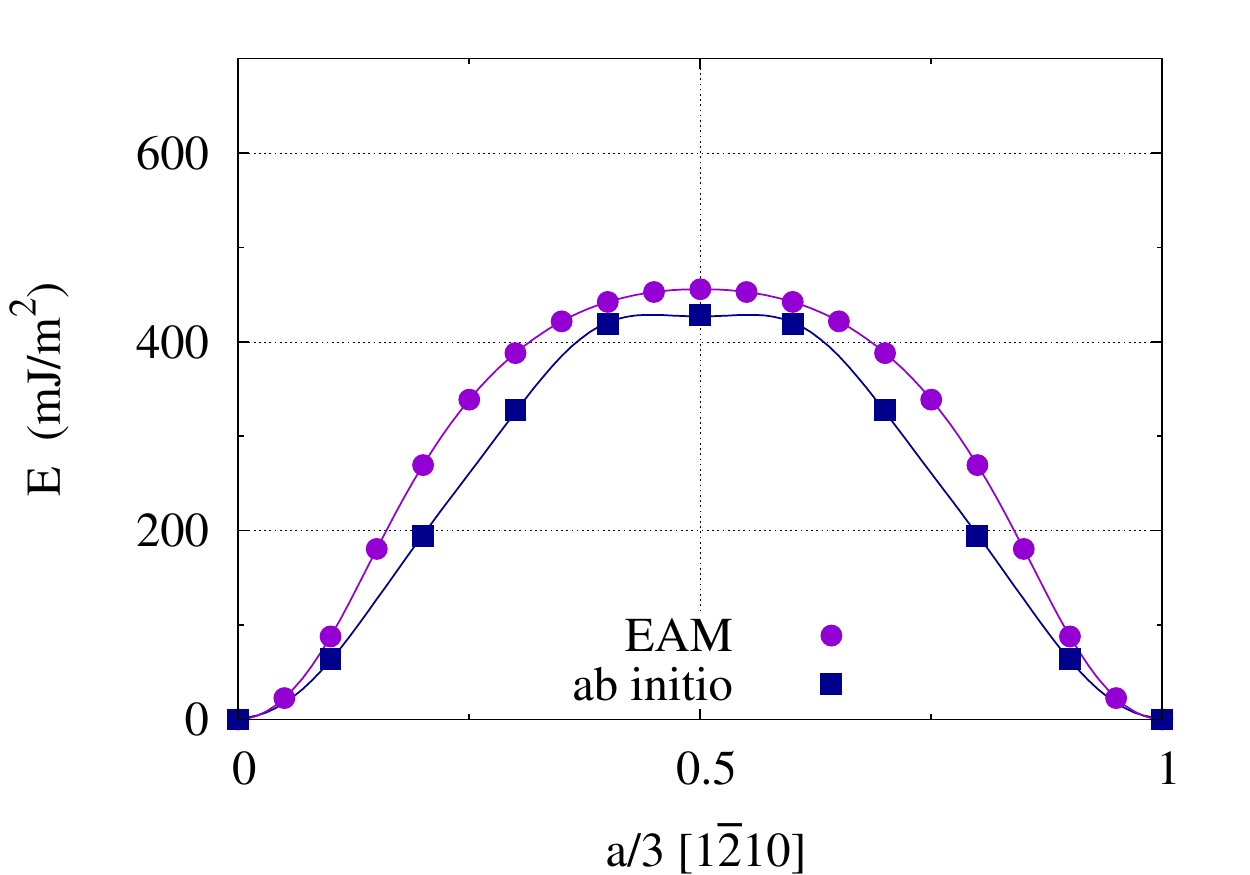}}
\caption{Generalized staking fault energy in the first order pyramidal plane $(10\bar{1}1)$ along the $[1\bar{2}10]$ direction calculated with ab initio and with the EAM potential.
The crystal is sheared either (a) in the loose plane $\pi_{1L}$, 
or (b) in the dense plane $\pi_{1D}$. (color online)}
\label{gammaLine}
\end{center}
\end{figure}

To compare both pyramidal planes, we plot in Fig. \ref{gammaLine} the generalized stacking fault energy only along the $[1\bar{2}10]$ direction for both planes. The energy obtained with ab initio calculations is higher than with the EAM potential in the case of the $\pi_{1L}$ plane, while it is nearly the same for the $\pi_{1D}$ plane. 
According to the EAM potential, the energy cost to shear the crystal along the $[1\bar{2}10]$ direction in a pyramidal plane is almost the same for both $\pi_{1L}$ and $\pi_{1D}$ planes. Ab initio calculations, however, predict that it is easier to shear in a $\pi_{1D}$ plane than in a $\pi_{1L}$ plane.
Our work therefore shows that both  $\pi_{1L}$ and $\pi_{1D}$ pyramidal planes may need to be considered when studying stacking faults in the first order pyramidal plane. This contrasts with the previous stacking fault calculations in hcp materials where only the $\pi_{1L}$ plane was considered \cite{Poty2011, Ghazisaeidi2012, Shin2012}.

Considering both Figs. \ref{projections} and \ref{gammaLine}, no energy minimum is found along the $[1\bar{2}10]$ direction, or in its immediate vicinity.
This is true for the ab initio calculations and the EAM potential. One should notice that the present $\gamma$-surfaces were obtained by relaxing the atoms only perpendicularly to the fault plane. A full atomic relaxation, however, allows for some atomic shuffling and reveals an energy minimum that corresponds to a metastable stacking fault in the $\pi_{1D}$ plane. The corresponding fault vector sketched by red arrow in Fig. \ref{projections}(b), is $\vec{f}= 1/2 \ \vec{a} + \vec{b_e}$, where $\vec{a} = 1/3\ [1\bar{2}10]$ and $\vec{b_e}$ is a  component orthogonal to $\vec{a}$ to be detailed below. This minimum is obtained with both ab initio calculations and the EAM potential.
On the other hand, no relevant minimum could be found for the $\pi_{1L}$ plane, even with full atomic relaxations.

\begin{figure}[!tbh]
\begin{center}
 \subfigure {\includegraphics[width=0.5 \linewidth]{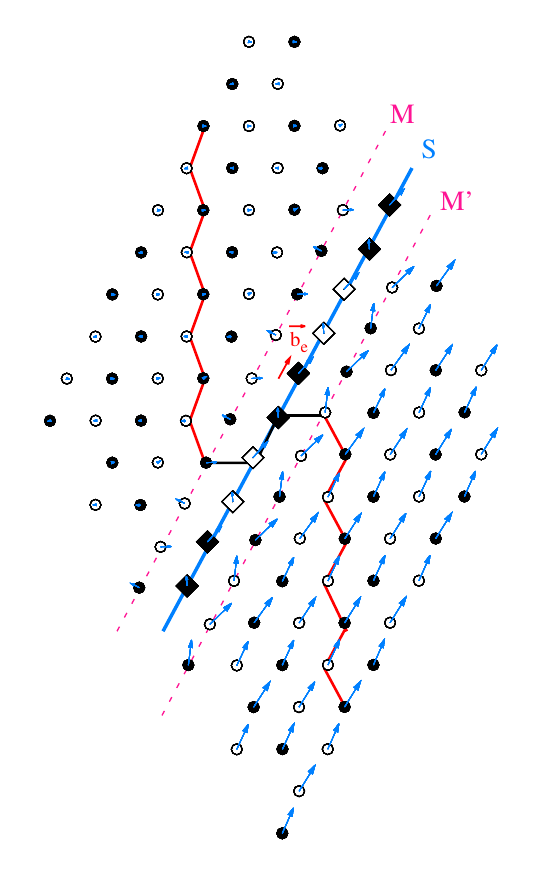}}
 \subfigure {\includegraphics[width=0.2 \linewidth]{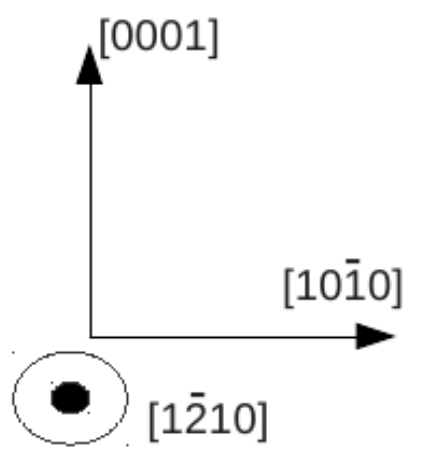}}
\caption{Atomic structure of the metastable pyramidal stacking fault: The displacement in the shearing $\langle a \rangle = [1\bar{2}10]$ direction is shown  by the projection of atoms in this direction, where atoms are sketched by black and white symbols depending on their position at half or full $\langle a \rangle$ vector respectively. The blue arrows show atom displacements perpendicular to the shearing direction in the $(1\bar{2}10)$ plane with a magnification factor of 3. The metastable fault corresponds to a pyramidal $\lbrace 10\bar{1}1 \rbrace$ twin bordered by two mirror planes M and M' sketched by pink dashed lines. The blue line S corresponds to the shearing plane. The mirror planes symmetry is highlighted by a broken line corresponding to the corrugated prismatic plane, in red in the parent hcp crystal and in black in the twinned crystal. Atoms with a neighborhood corresponding to the twinned crystal are sketched by diamonds, while the circles corresponds to atoms in the parent crystal. (color online)}
\label{stFault}
\end{center}
\end{figure}

The atomic structure of the metastable stacking fault is shown in Fig. \ref{stFault}. The displacement map shows that in the shearing direction $[1\bar{2}10]$, the atoms below the shearing plane S have their color switched, which means that they have been displaced by $a/2$. The shearing is thus perfectly localized in the S plane. 
Perpendicularly to the shearing direction, the blue arrows show a displacement of all the atoms below the S plane following one same vector $\vec{b_e}$, which corresponds to the orthogonal component of the fault vector, as well as a shuffling of the atoms which is extended to several planes at both sides of the S plane. 
 This shuffling explains why the metastable fault did not appear on the $\gamma$-surfaces of Fig. \ref{projections}, where only atomic relaxation perpendicularly to the fault plane was allowed.
 
Analysis of this metastable core structure reveals that it corresponds to an elementary two-layer pyramidal $\lbrace 10\bar{1}1 \rbrace$ twin \cite{Chaari2014} bordered by the M and M' mirror planes, as illustrated by the broken line on Fig. \ref{stFault}. We also looked at the positions of the first nearest neighbors for each atom and compared the obtained pattern with the ones existing in a perfect hcp structure, both for the parent and the twinned lattices.  We thus managed to characterize whether an atom belongs to the parent or the twinned hcp lattice. On figure \ref{stFault}, atoms plotted with diamonds corresponds to the atoms belonging to the twin layer. 

 The twin is produced by the glide of a two-layer disconnection with a Burgers vector that corresponds to the fault vector $\vec{f}= 1/2 \ \vec{a} + \vec{b_e}$ where the edge component of the disconnection is defined by $f_e= a(4\gamma^2-9)/2\sqrt{3+4\gamma^2}$ ($\gamma$ is the $c/a$ ratio) \cite{Serra1991, Wang2011}. Two layer disconnections are well-known to be stable on pyramidal twins \cite{Wang2011,Rodney2013, Bursill1995, Li2009, Pond1995, Serra1991, Wang2012}, but the stability of the corresponding two-layer twin was so far unknown.
 
The stacking fault energy deduced from ab initio calculations is $\Delta E= 163$ mJ\,m$^{-2}$.
It is lower than the energy of the prismatic stacking fault ($\Delta E = 211$ mJ\,m$^{-2}$) \cite{clouet2012}.
Compared to the ab initio value, the EAM potential overestimates the pyramidal stacking fault energy ($\Delta E= 243$ mJ\,m$^{-2}$).
This leads to a higher energy than for the prismatic fault ($\Delta E= 135$ mJ\,m$^{-2}$)

\section{Peierls barrier in the pyramidal plane}

Considering the results for the generalized stacking faults, we conclude that it is important to include both pyramidal $\pi_{1L}$ and $\pi_{1D}$ planes in our study.  We thus investigate in this part the glide of an $\langle a \rangle$ screw dislocation in first order pyramidal planes, for both $\pi_{1L}$ and $\pi_{1D}$ planes.  

\subsection{EAM}
 
Starting from a dislocation in its equilibrium configuration, i.e. initially spread in a prismatic plane (Fig. \ref{disloCore}(a)), we calculated the energy encountered by the dislocation to overcome a Peierls valley in the pyramidal plane, moving to a final equilibrium state where the dislocation spreads in the next prismatic plane. We used the NEB method to calculate the Peierls barrier in both pyramidal $\pi_{1L}$ and $\pi_{1D}$ planes with the EAM potential. The initial path of the dislocation is obtained by a linear interpolation between the initial and the final states with the cut created by the dislocation glide localized in the chosen first order pyramidal plane. 
 
\begin{figure}[!tbh]
\begin{center}
  \includegraphics[width=0.8 \linewidth]{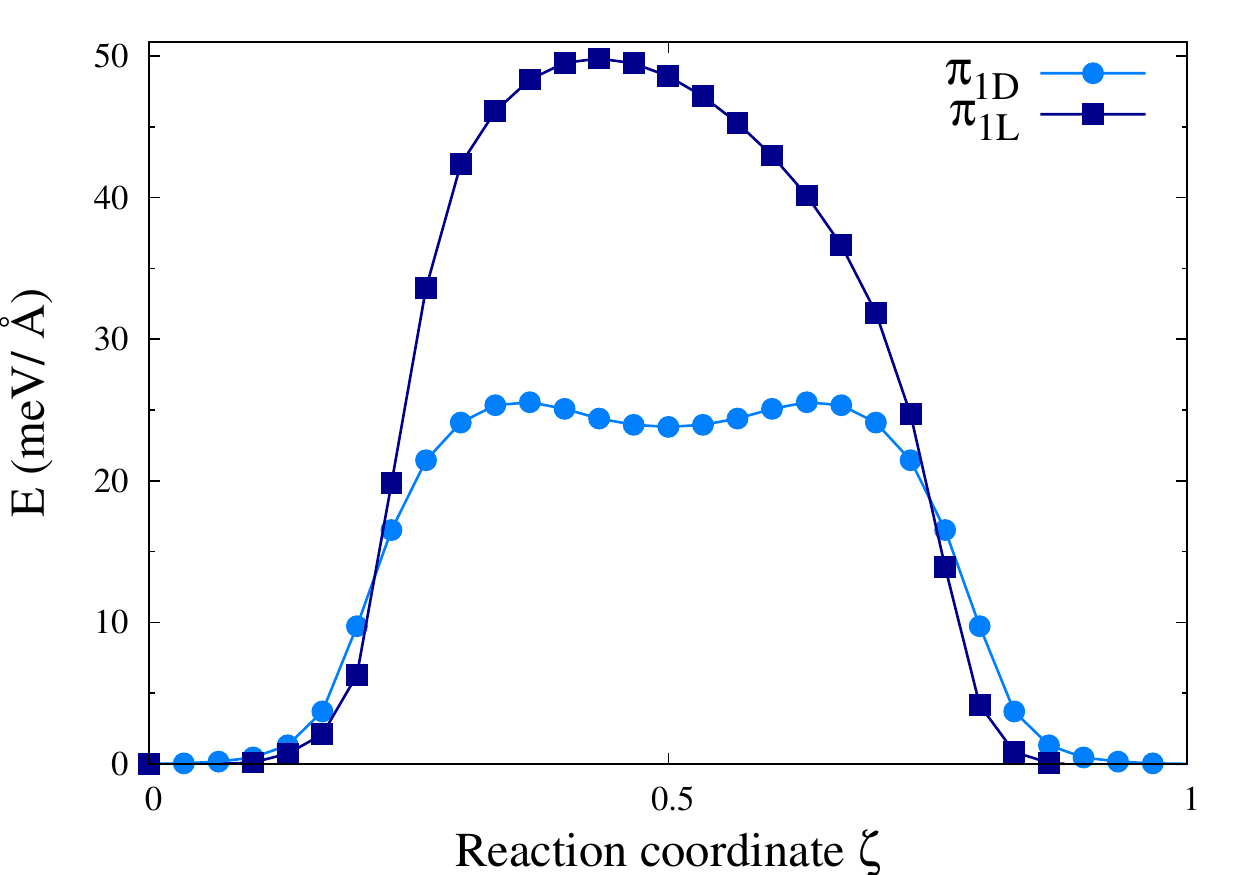}
\caption{Energy barrier encountered by a screw dislocation dissociated in a prismatic plane and gliding in pyramidal $\pi_{1L}$ and $\pi_{1D}$ planes calculated with the EAM potential. (color online)}
\label{barriereEAM}
\end{center}
\end{figure}

The results are shown in Fig. \ref{barriereEAM}. The energy barrier in the $\pi_{1L}$ plane is twice higher than in the $\pi_{1D}$ plane. 
Thus, according to the EAM potential, it is easier for the dislocation to glide in the $\pi_{1D}$ plane than in the $\pi_{1L}$ plane.
  
The energy path obtained in the $\pi_{1D}$ plane shows a local minimum at halfway across the migration. This minimum corresponds to an intermediate metastable configuration of the screw dislocation (Fig. \ref{disloCore}(b)), to be described in more details below. 

\subsection{Ab initio}

We only consider dislocation glide in the $\pi_{1D}$ plane for the ab initio calculations. 
This is motivated by the ab initio results for the generalized stacking faults, showing that the $\pi_{1D}$ plane is easier to shear in the $[1\bar{2}10]$ direction than the $\pi_{1L}$ plane (Fig. \ref{gammaLine}), as well as by the Peierls barrier obtained with the EAM potential, showing that dislocation glide in the $\pi_{1D}$ plane costs less energy than in the $\pi_{1L}$ plane (Fig. \ref{barriereEAM}).

\begin{figure}[!tbh]
\begin{center}
  \includegraphics[width=0.8 \linewidth]{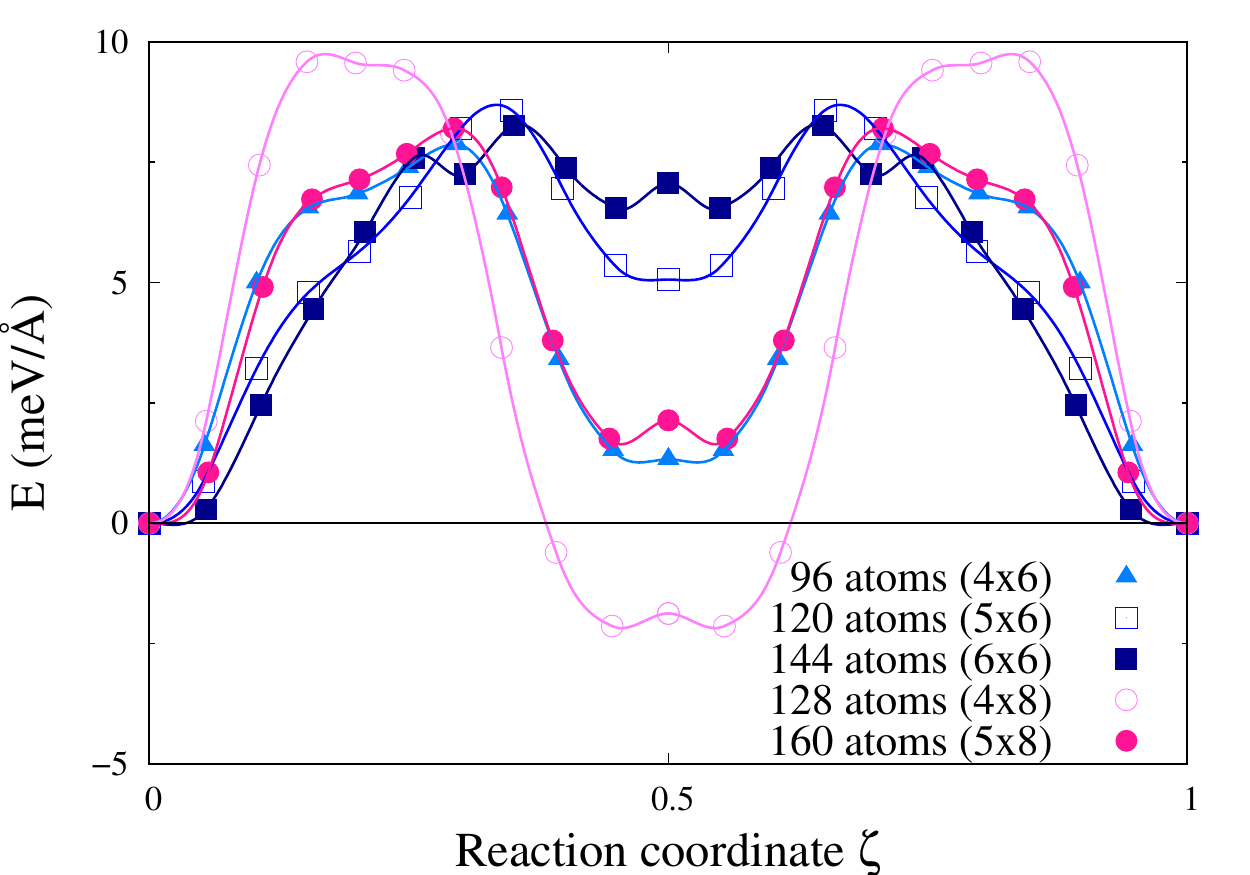}
\caption{Energy barrier encountered by a screw dislocation dissociated in a prismatic plane when gliding in a pyramidal $\pi_{1D}$ plane. Ab initio calculations are performed for different simulation cell sizes $n \times m$. (color online)}
\label{barriereAbinitio}
\end{center}
\end{figure}

Ab initio calculations of the Peierls barrier in the $\pi_{1D}$ plane were performed for different simulation cell sizes.
The minimum energy paths obtained are illustrated in Fig. \ref{barriereAbinitio}. They all show a local minimum  halfway across the migration, in agreement with the EAM results. This minimum corresponds to the same intermediate metastable configuration of the screw dislocation as found with the EAM potential.
 
 The ab initio energy barrier is twice lower than with the EAM potential. The difference in energy between ab initio and EAM is related to the fact that the Mendelev potential overestimates the energy of the pyramidal metastable stacking fault. As a consequence this empirical potential leads to a higher Peierls barrier in the pyramidal plane.

\section{Metastable configuration of the screw dislocation in zirconium}

Both ab initio calculations and the EAM potential showed that pyramidal glide involves an intermediate  metastable configuration of the screw dislocation appearing halfway across the migration. In the following, a detailed description of this metastable configuration is proposed.

\subsection{Core structure}

\begin{figure}[!tbh]
\begin{center}
\subfigure [EAM: stable] {\includegraphics[width=0.45\linewidth]{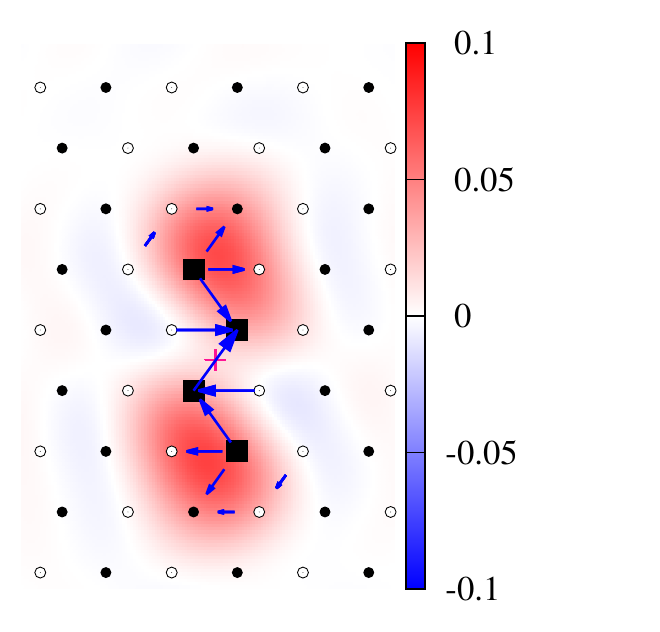}}
\subfigure [EAM: metastable] {\includegraphics[width=0.35\linewidth]{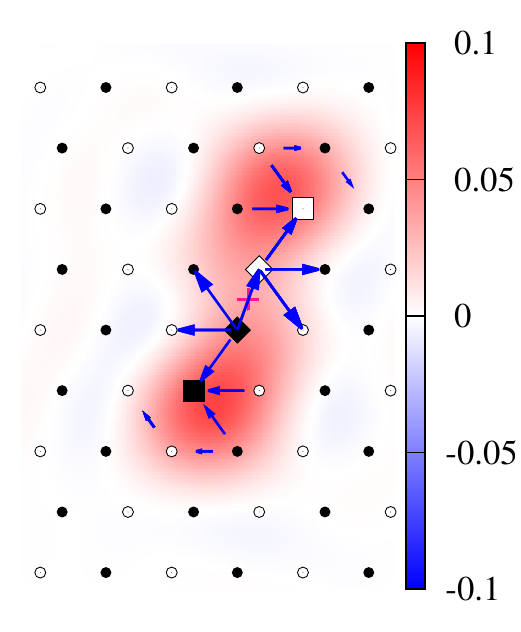}}
\hfill
\subfigure [Ab initio: stable] {\includegraphics[width=0.45\linewidth]{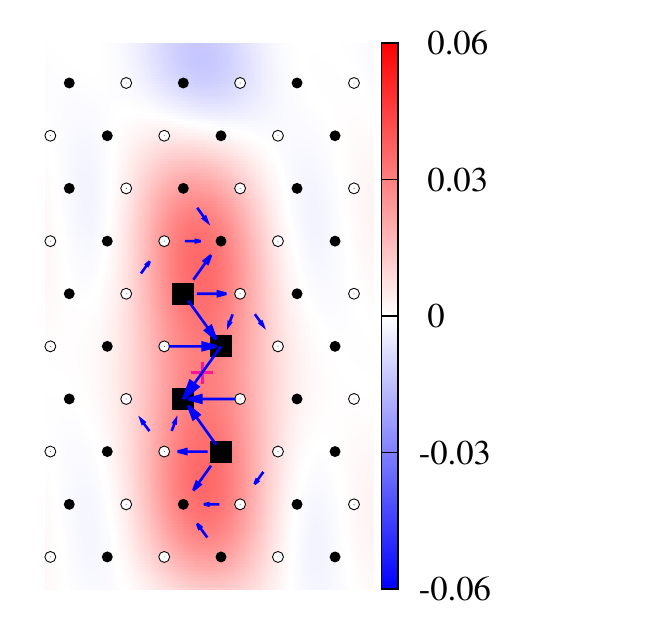}}
\subfigure [Ab initio: meta\-stable] {\includegraphics[width=0.35\linewidth]{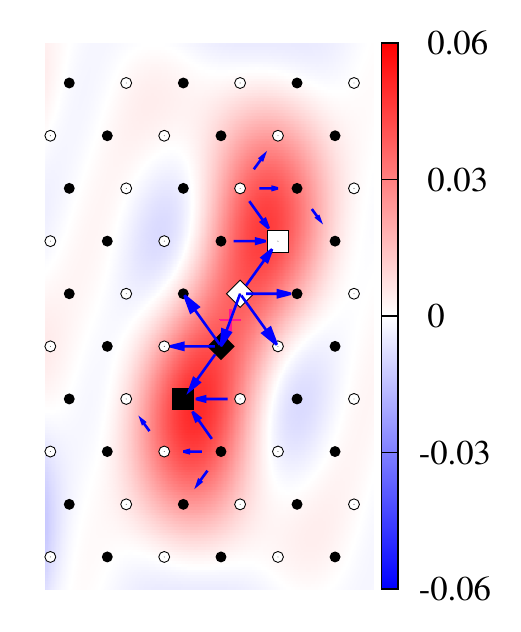}}
\hfill
\caption{Differential displacement map of the screw dislocation in its equilibrium and metastable configuration deduced from ab initio calculations and EAM potential. The black and white symbols corresponds to atoms depending on their position along the $\langle a \rangle = [1\bar{2}10]$ direction in the perfect crystal, while the blue arrows corresponds to the differential displacement between two columns of atoms in the $\langle a \rangle$ direction after relaxation with dislocations. The pink cross indicates the dislocation center obtained by symmetry. The screw component of the Nye tensor distribution is superimposed. Squares correspond to atoms in a prismatic fault neighborhood while diamonds correspond to atoms in the pyramidal twin neighborhood. (color online)}
\label{disloCore}
\end{center}
\end{figure}

Fig. \ref{disloCore} shows the core structure of the two possible configurations obtained for the $\langle a \rangle$ screw dislocation with ab initio calculations and with the EAM potential.
The differential displacement maps have been superimposed to the Nye tensor distribution calculated following the method of Hartley and Mishin \cite{Hartley2005}. Only the screw component of the Nye tensors is plotted in each structure.  
This Burgers vector density is deduced from the position variation of the nearest neighbors for each corresponding atom. In a perfect hcp structure each atom has twelve nearest neighbors forming a defined pattern. However in a faulted structure, this number of nearest neighbors can be different with different corresponding patterns. In the figure, atoms belonging to a pattern that corresponds to a prismatic stacking fault have been plotted as squares while those belonging to a pattern that corresponds to the pyramidal twin described before are plotted as diamonds. 
We can see through Fig. \ref{disloCore} that ab initio calculations and EAM potential results are in good agreement with the same metastable core obtained in both cases. 

 The equilibrium configuration of the dislocation (Fig. \ref{disloCore}(a) and (c)) shows a spread in the prismatic plane in agreement with the literature \cite{clouet2012, Bacon2002}. The dissociation of the dislocation into two partials is illustrated by two local extrema in the Nye tensor distribution and the prismatic stacking fault in the core is highlighted by the squares showing the atoms involved in the fault.  

Analyzing the displacement maps of the metastable core structure (Fig. \ref{disloCore}(b) and (d)), we show that the dislocation is spread in three different crystallographic planes at the same time: in the core center, the dislocation lies in a pyramidal plane while it lies in two adjacent prismatic planes at the extremities. The two central atoms sketched by black and white diamonds (Fig. \ref{disloCore}(b) and (d)) witness of the presence of the pyramidal twin pattern, while the squares result from a local shearing in the prismatic plane. 

 The pyramidal spreading in the core center is explained by the metastable stacking fault evidenced above in the pyramidal $\pi_{1D}$ plane with a fault vector $\vec{f}= 1/2 \ \vec{a} + \vec{b_e}$. The central part of the core thus corresponds to an elementary two layer twin of a finite extension. Since the screw component of the fault vector, $1/2 \ \vec{a}$, is identical in both the prismatic and pyramidal faults, there is no discontinuity in the screw direction at the intersection between the faults. The pyramidal fault is thus bordered at its intersections with the prismatic fault by two edge disconnections of slip vectors $\pm \vec{b_e}$, while the prismatic faults end with screw partial dislocations with $1/2 \ \vec{a}$ Burgers vectors. This can be seen through the Nye tensor distribution plot in Fig. \ref{disloCore}(b) and (d) where two partial dislocations are distinguished in two neighboring prismatic planes. 

The metastable core may thus be described as two partial dislocations spread in two adjacent prismatic planes, separated by two prismatic stacking faults and a pyramidal nanotwin in-between \cite{Chaari2014}. The prismatic stacking faults are linked to the nanotwin by stair-rods forming a dipole of disconnections ($\vec{b_e}$).   
The corresponding decomposition of the total Burgers vector is 
\begin{equation*}
\frac{1}{3} \ a[1\bar{2}10] \to \frac{1}{6} \ a [1\bar{2}10]
		+ \frac{b_e}{\sqrt{3+4\gamma^2}}[10\bar{1}2]
		- \frac{b_e}{\sqrt{3+4\gamma^2}}[10\bar{1}2]
		+ \frac{1}{6} \ a [1\bar{2}10], 
\end{equation*}
where $\gamma$ is the $c/a$ ratio.

\subsection{Core energy}

\begin{figure}[!tbh]
\begin{center}
\subfigure [EAM] {\includegraphics[width=0.49 \linewidth]{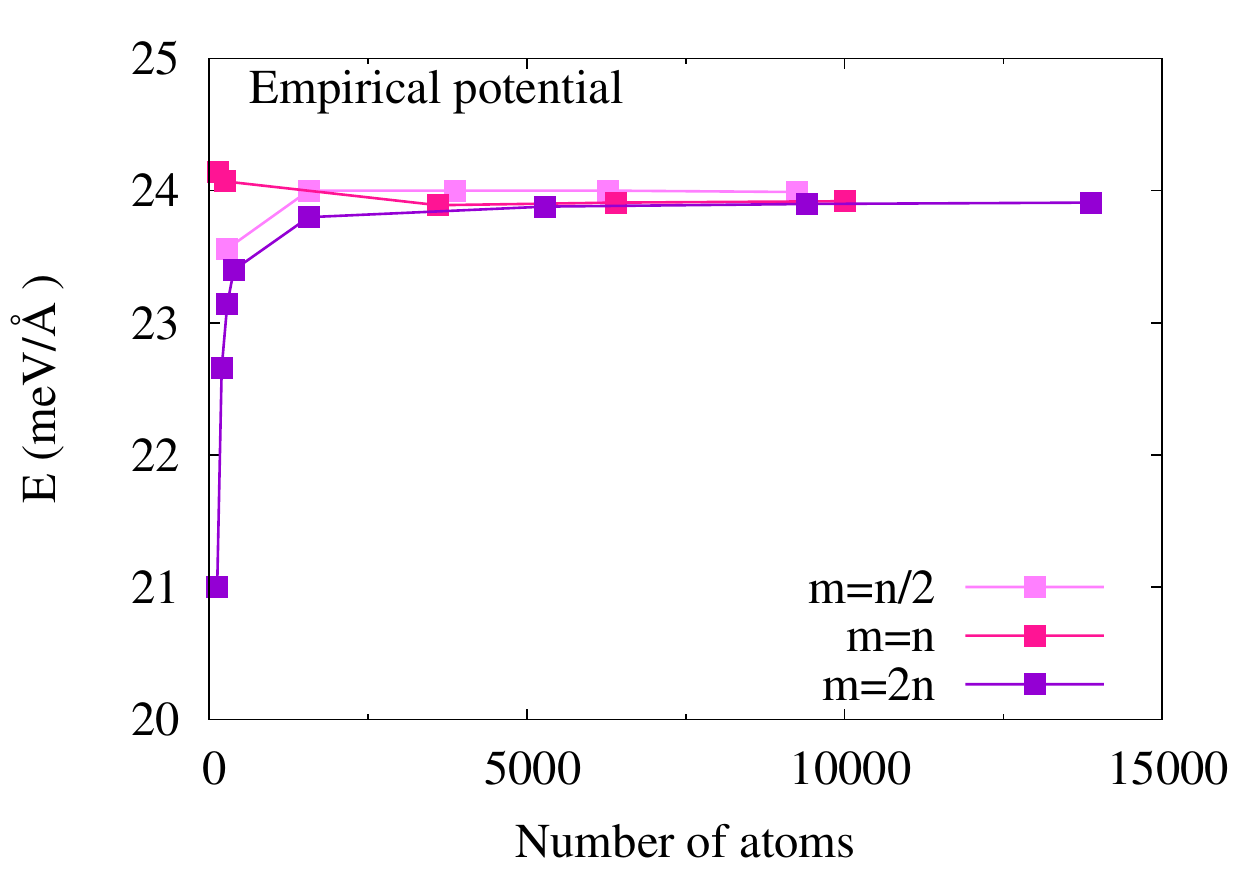}}
\subfigure [Ab initio] {\includegraphics[width=0.49 \linewidth]{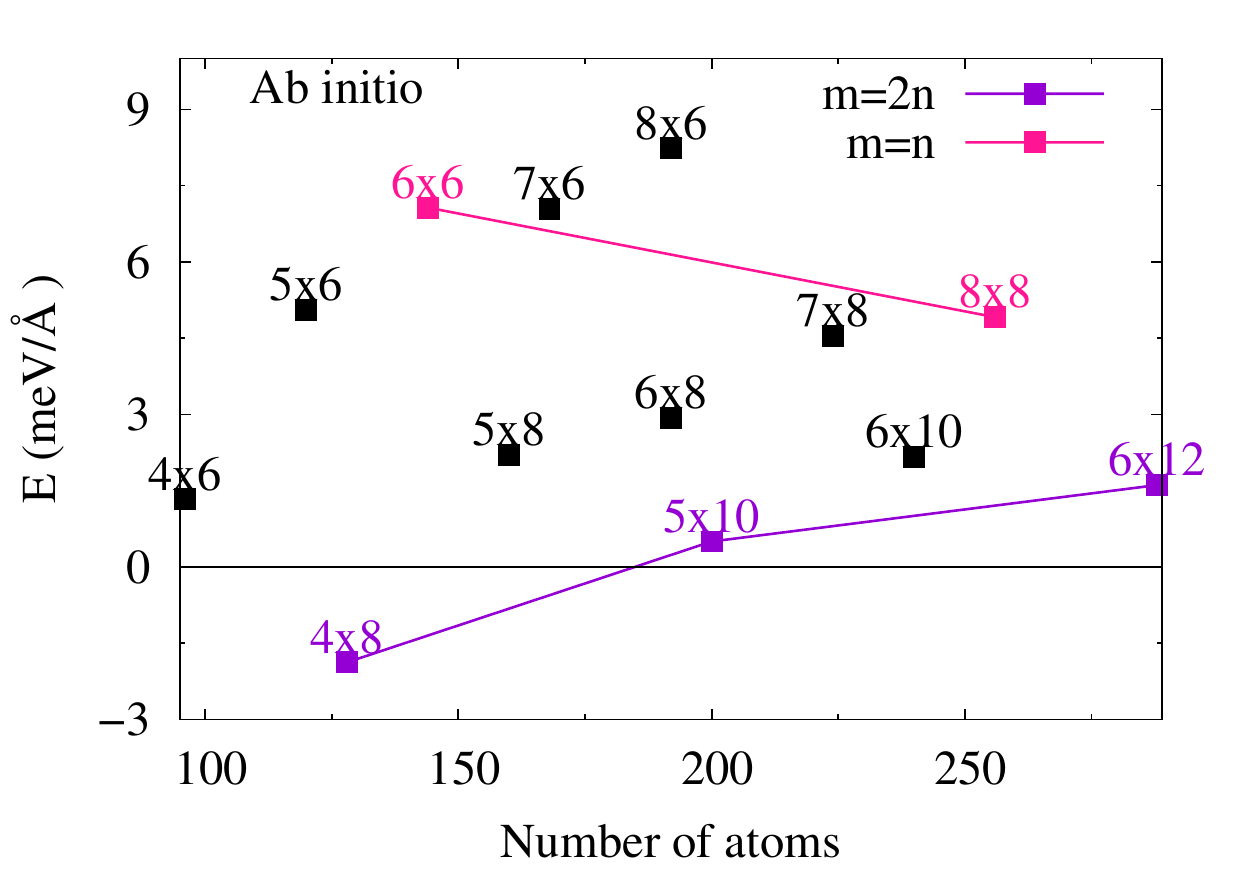}}
\caption{Difference between the energy of the metastable configuration and the equilibrium configuration of the screw dislocation for different simulation cell sizes $n\times m$ obtained with the EAM potential on the left and with ab initio calculations on the right. (color online)}
\label{interEnergy}
\end{center}
\end{figure}

Fig. \ref{interEnergy} summarizes, for different cell sizes, the excess energy of the metastable configuration, with respect to the energy of the equilibrium configuration fully dissociated in the prismatic plane. 
With the EAM potential, this excess energy is always positive, confirming that the metastable core is less favorable than the prismatic configuration. 
Convergence of the results is obtained for simulation cells containing more than 2000 atoms (Fig. \ref{interEnergy}(a)), with an excess energy of $\Delta E= 24$\,meV/\AA. 
Different simulation cell shapes lead to different convergence behaviors. At small sizes, an upper bound of the converged excess energy is obtained with shapes defined by $m=n$, whereas a lower bound is obtained with $m=2n$.

Ab initio calculations lead to a lower excess energy (Fig. \ref{interEnergy}(b)). As a consequence, stability inversion was observed for very small simulation cells ($4\times8$ cell containing only 128 atoms). However, larger simulation cells confirm that the configuration partially spread in the pyramidal plane is metastable.
Since ab initio calculations are more expensive and limited to few hundred atoms, it was not possible to reach a converged value of the energy. However, the same dependence of the convergence rate with the cell shape was observed with the ab initio calculations and with the EAM potential. An upper limit of the energy is thus given by the $m=n$ cells and a lower limit by the $m=2n$ cells. Our ab initio calculations therefore lead to an excess energy $\Delta E= 3.2 \pm 1.6$\,meV/\AA. 

\section{Conclusion}

In the present work, based on generalized stacking fault calculations in the first order pyramidal plane, we demonstrated that shearing along the $[1\bar{2}10]$ direction inside a dense pyramidal plane costs  less energy than between two corrugated planes. In addition, calculations showed a metastable stacking fault in the pyramidal plane, which corresponds to an elementary pyramidal two-layer twin. This metastable stacking fault is at the origin of the new metastable core configuration of the $\langle a \rangle$ screw dislocation in zirconium, which appears halfway across the migration path when the dislocation glides in a pyramidal plane.  This metastable configuration presents an unusual core structure with an incipient two-layer twin in its center.

We conclude that there are two possible configurations of the screw dislocation in zirconium. The one with the lower core energy, is dissociated in the prismatic plane and responsible for the easy glide in this plane \cite{clouet2012}. The second configuration is metastable and appears during pyramidal and basal slips \cite{Chaari2014}. 
The results show a good agreement between ab initio calculations and the Mendelev empirical potential since, qualitatively, both lead to the same glide mechanism and metastable core. This work is a first step toward  understanding cross-slip in zirconium, showing a new and unexpected relation between dislocation glide and twinning, two essential motors for hcp plasticity. 

\section{Acknowledgments}
	This work was performed using HPC resources from GENCI-CINES, -CCRT and -IDRIS
  (Grants 2013-096847).
  The authors also acknowledge PRACE for awarding access to the Marenostrum resources 
  based in Barcelona Supercomputing Center (project DIMAIM).

\bibliographystyle{unsrt}
\bibliography{article}

\end{document}